\RequirePackage{amsmath}
\documentclass[10pt]{iopart}
\usepackage{amsmath}
\usepackage{amsfonts}
\usepackage{amssymb}
\usepackage{graphicx}
\newcommand{\hx}{\hat{x}}
\newcommand{\hy}{\hat{y}}
\newcommand{\hr}{\hat{r}}
\newcommand{\hphi}{\hat{\varphi}}
\newcommand{\erf}[1]{\textrm{Erf}\left[#1\right]}
\newcommand{\erfc}[1]{\textrm{Erfc}\left[#1\right]}
\newcommand{\erfi}[1]{\textrm{Erfi}\left[#1\right]}
\newcommand{\pFq}[5]{{_{#1}F_{#2}}\left(#3,#4,#5\right)}

\begin{document}

\title{Non-commutativity in polar coordinates}
\author{James P. Edwards}
\address{Mathematical Sciences Department, University of Bath, Claverton Down, Bath, BA2 7AY, UK.}
\ead{jpe28@bath.ac.uk}

\begin{abstract}
We reconsider the fundamental commutation relations for non-commutative $\mathbb{R}^{2}$ described in polar coordinates with non-commutativity parameter $\theta$. Previous analysis found that the natural transition from Cartesian coordinates to polars led to a representation of $\left[\hr, \hphi\right]$ as an everywhere diverging series. We compute the Borel resummation of this series, showing that it can subsequently be extended throughout parameter space and hence provide an interpretation of this commutator. Our analysis provides a complete solution for arbitrary $r$ and $\theta$ that reproduces the earlier calculations at lowest order. We compare our results to previous literature in the (pseudo-)commuting limit, finding a surprising spatial dependence for the coordinate commutator when $\theta \gg r^{2}$. We raise some questions for future study in light of this progress.
\end{abstract}

\maketitle

\ioptwocol

\section{Introduction}
Non-commutative spaces have become of recent interest for a wide variety of topics in high energy physics. Such ideas appear naturally in some applications of string theory \cite{String1, String2}, in particular for string propagation in a background electromagnetic field or Kalb-Ramond form, and are postulated in some models of quantum gravity \cite{gravity1, gravity2}. A common starting point in such a setting is the introduction of the fundamental commutator between spatial coordinates 
\begin{equation}
	\left[\hx_{i}, \hx_{j}\right] = i\theta_{ij}
	\label{CommC}
\end{equation}
for constant non-commutativity parameters assembled into a real, skew-symmetric matrix $\theta_{ij}$. Much progress has been made in understanding various aspects of these non-commutative spaces \cite{Review} and their application to quantum field theory \cite{qft1, qft2, WL, qft3}, where the effects of UV / IR mixing have played an important r\^{o}le. There is also interest in generalisations of (\ref{CommC}), perhaps most famously to describe the ``fuzzy sphere'' \cite{sphere, Coherent}. In this note we revisit a two dimensional non-commutative space described by polar coordinates. 

One facet of non-commutative geometry that has not often been discussed is its application in non-Cartesian coordinates. However, in \cite{btz}, as a refinement of \cite{btz2}, the authors considered black holes in a non-commutative version of $AdS_{3}$, using the polar variables, $\hr, \hphi$, to describe the spatial coordinates. One outstanding issue was that the transition to the basic commutator in polar coordinates, $\left[\hat{r}, \hat{\varphi}\right]$, was not justified, which in turn led Iskauskas to investigate how it might be deduced \cite{andy}. The result was somewhat surprising; as we will briefly recapitulate below, starting from (\ref{CommC}) it was possible to find a power series expansion for the polar commutator, which unfortunately suffered from a vanishing radius of convergence. Consequently, the conclusion of \cite{andy} was to treat $\hr^{2}$ and $\hphi$ as the natural non-commuting variables -- the power series for their commutator was found to converge everywhere and was evaluated to give $\left[\hr^{2}, \hphi\right] = 2i\theta$, in agreement with the commutator used for simplicity\footnote{It is easy to check that this is compatible with the commutator $\left[\hr, \hphi\right] = \frac{i \theta}{r}$ postulated in \cite{btz}.} in \cite{btz}.

We briefly return here to the problematic expansion for $\left[\hr, \hphi\right]$ to show that it is possible to give some meaning to this series. As we will show, it turns out to have a Borel resummation which will allow us to give an interpretation to the commutator. In fact we will find that the result can be used to define the commutator everywhere on the plane and for all values of the non-commutativity parameter $\theta$. This is a physically significant problem since with three spatial dimensions (assuming a flat, non-compact manifold) a suitable rotation takes the space described by (\ref{CommC}) into the product of a two-dimensional non-commuting space and the (commuting) real line, wherein symmetry considerations may make polar coordinates a natural description of the non-commutative plane. In the following section we recap the calculation of the polar commutator in \cite{andy} before presenting the Borel resummation of the resulting series. We then investigate its compatibility with the commutators previously employed in the literature and explore other limiting cases. It is hoped that this might clarify the differences between the choices for the fundamental commutator in polar coordinates, which is an important point for further understanding of non-commutativity on general spaces. 

\section{Non-commutative polar coordinates}
We quickly outline the calculation carried out in \cite{andy}. Starting with $\left[\hx, \hy\right] = i\theta_{12} \equiv i\theta$ we relate this (at least formally) to polar variables via the familiar relationship $\hr = \sqrt{\hx^{2} + \hy^{2}} > 0$ and $\hphi = \arctan\left(\frac{\hy}{\hx}\right)$. To avoid the ambiguity of the variable ordering in these identifications, it proved advantageous to use the well-known correspondence between these non-commuting operators and the space of functions on $\mathbb{R}^{D}$ under a non-commuting product. Such a map is exhibited by the Weyl symbol
\begin{flalign}
	\hat{\mathcal{O}}\left[f\right] &= \int d^{D}x\, f(x)\hat{\Delta}\left(x\right); \nonumber \\
	 \hat{\Delta}(x) &= \int \frac{d^{D}k}{\left(2\pi\right)^{D}} e^{i k \cdot \hx} e^{-i k \cdot x}.
\end{flalign}
From this map the Moyal $\star$-product \cite{Moyal, Moyal2} is defined to deduce the position space representation of the product of two Weyl operators
\begin{flalign}
	\hat{\mathcal{O}}\left[f\right]\hat{\mathcal{O}}\left[g\right] &= \hat{\mathcal{O}}\left[f\star g\right]; \nonumber \\
	f(x)\star g(x) &= f(x) \exp{\left(\frac{i}{2}\overset{\leftarrow}{\partial_{i}}\,\theta_{ij}\!\stackrel{\rightarrow}{\partial_{j}}\right)}g(x)
	\label{star},
\end{flalign}
which acts as a deformation of the usual algebra of functions on $\mathbb{R}^{D}$ \cite{Lizzi, Product}. We use the customary notation that the derivatives act on the functions in the directions of their overhead arrows.

This was used in \cite{andy} to overcome the ordering ambiguities in the transition to polar coordinates by instead considering the commutator of the real functions $f(x, y) = r = \sqrt{x^{2} + y^{2}}$ and $g(x, y) = \phi = \arctan\left(\frac{y}{x}\right)$ under the $\star$-product:
\begin{equation}
	\left[r, \phi\right]_{\star} \equiv r \star \phi - \phi \star r.
	\label{CommS}
\end{equation}
By formally expanding the exponential in (\ref{star}) it was possible to arrive at the power series solution\footnote{The restriction $r > 0$ is important for (\ref{series}) which holds up to dependence on $\delta^{2}(r)$ and its derivatives.}
\begin{equation}
	\left[r, \phi\right]_{\star} = \frac{i\theta}{r} \sum_{n=0}^{\infty} \frac{\left(4n\right)!}{\left(2n+1\right)!} \left(\frac{\theta}{4r^{2}}\right)^{2n}.
	\label{series}
\end{equation}
It is simple to verify that the radius of convergence of this series is zero, from which it is tempting to abandon the commutator as a sensible object. In the next section, however, we will calculate its Borel resummation to enlarge its radius of convergence, the result of which will be possible to continue to all positive values of $\theta$ and will be valid for all values of $r$.

\subsection{The resummation of the series}
In general, consider a power series in the (typically complex) variable $z$ with radius of convergence $R < \infty$ defined by	$A(z) = \sum_{n=1}^{\infty} a_{n}z^{n}$. In many cases one may sensibly enlarge\footnote{In the sense that when the original series converges then the result of the Borel resummation converges to the same value.} the radius of convergence. This procedure is motivated by using the Gamma function to express $z^{n}$ as an integral and consequently swapping the order of integration and summation to arrive at the Borel sum
\begin{equation}
	A_{B}(z) = \int_{0}^{\infty} dt \, \mathcal{B}_{a}(t) e^{-t / z}
	\label{Bo}
\end{equation}
where $\mathcal{B}_{a}(t)$ is the Borel transform of the series
\begin{equation}
	\mathcal{B}_{a}(t) = \sum_{n=1}^{\infty} \frac{a_{n} t^{n-1}}{\left(n-1\right)!}.
	\label{BoT}
\end{equation}
The point is that the series (\ref{BoT}) may have better convergence properties since its terms grow more slowly than those of the original series by a factor of $\left(n-1\right)!$ -- one may also hope that this improvement is transferred to the resulting expression in (\ref{Bo}) and that it may then prove possible that it be continued to an even wider region for $z$. 

We adapt this technique to the series in (\ref{series}), defining first $z = \frac{\theta}{4r^{2}}$ and stripping the first term from the sum. Then for $n \geqslant 1$ we may write $z^{2n} = \frac{1}{\left(2n-1\right)!}\int_{0}^{\infty}dt\, t^{2n-1} e^{-t / z}$ which leads to the Borel transform
\begin{equation}
	\mathcal{B}(t) =  \sum_{n = 1}^{\infty}\frac{\left(4n\right)! t^{2n-1}}{\left(2n-1\right)! \left(2n+1\right)!}\,.
\end{equation}
This series in the variable $t$ has radius of convergence equal to $\frac{1}{4}$ which is indeed an improvement on the original series in (\ref{series}). Furthermore, we have evaluated this sum in closed form,
\begin{equation}
	\mathcal{B}(t) = -\left(\frac{2t -1}{4t^{2}\sqrt{1 - 4t}} + \frac{2t+1}{4t^{2}\sqrt{1 + 4t}}\right),
\end{equation}
so that the resummation of the commutator can be expressed as
\begin{flalign}
	&\left[r, \phi\right]_{\star B}(\theta) = \frac{i\theta}{r}\bigg[1 -\nonumber \\
	 &\qquad \int_{0}^{\infty}dt \left(\frac{2t -1}{4t^{2}\sqrt{1 - 4t}} + \frac{2t+1}{4t^{2}\sqrt{1 + 4t}}\right) e^{-t / z} \bigg],
	\label{seriesB}
\end{flalign}
whose required integral we now look to compute.

One notes the presence of the divergence at $t = \frac{1}{4}$, as can be expected given the radius of convergence of $\mathcal{B}(t)$. However, this poses no problem in (\ref{seriesB}) as it is too soft to damage the convergence of the integration\footnote{The required branch cut forces us to define precisely the integration over the parameter $t$, which we take with infinitesimal positive imaginary part; we would encounter the usual Stokes phenomenon \cite{Stokes} were we to deform the integration contour to run just below the real axis.}. Moreover, the perceived divergence of the integrand as $t \rightarrow 0$ is easily seen to be artificial by verifying that the Taylor expansion of the Borel transform about $t = 0$ takes the form
\begin{equation}
	\mathcal{B}(t) =4t + \mathcal{O}\left(t^{3}\right).
\end{equation}
Indeed, considering also the behaviour of the integrand as $t \rightarrow \infty$ we see that the integral converges for all $\Re(z) \geqslant 0$ and as such provides us with a means of interpreting the commutator after all. To check that such an interpretation is compatible with previous results, we may already see that in the limit of vanishing $\theta$ (representing a commuting space) or the (pseudo-commutative) limit $r \rightarrow \infty$ we have that $z \rightarrow 0$ and we recover the previously postulated commutation relation $\left[r, \phi\right]_{\star} = \frac{i\theta}{r}$. We are interested here in determining the corrections to this result implied by (\ref{seriesB}).

A simple observation is that the integration required to complete the Borel resummation in (\ref{Bo}) and (\ref{seriesB}) can be thought of as finding the Laplace transform of the Borel transform of the terms in the series, thought of as a function of $\frac{1}{z}$. Consequently we may use various well-known properties of the Laplace transform. In particular, for smooth functions $f(t)$ with Laplace transform $F(s) \equiv \mathcal{L}\left[f\right](s)$, we will require the result that $\lim_{s \rightarrow \infty} F(s) = 0$ and that $\mathcal{L}\left[t f(t)\right](s) = -F^{\prime}(s)$ which allows us to deduce the Laplace transforms of $t^{-1}f(t)$ and $t^{-2}f(t)$ up to constants
\begin{align}
	\mathcal{L}\left[t^{-1}f(t)\right](s) &\simeq -\int^{s} \!ds^{\prime} F(s^{\prime})  \nonumber \\
	 \mathcal{L}\left[t^{-2}f(t)\right](s) &\simeq \int^{s}\!\!\int^{s^{\prime}} \!ds^{\prime}ds^{\prime\prime} F(s^{\prime\prime}) \nonumber \\
	 &\simeq \int^{s}  \!ds^{\prime} \left(s - s^{\prime}\right)F(s^{\prime}).
	 \label{Laplace}
\end{align} 
We may use these properties to calculate -- at least up to such undetermined constants -- the Laplace transforms of the individual terms in (\ref{seriesB}) and then use their asymptotic behaviour to infer the remaining $s$-independent term when these parts are combined to a result that we know is well-defined (we identify $s =z^{-1}$). In this way we regulate the small-$t$ divergences of the individual terms under the integral (in the sense of introducing a cut-off, so that the undetermined constants may, in intermediate steps, be divergent in the limit that the regulator is removed). Our result can then be checked against a numerical evaluation of the integral to verify the analytic calculation is correct. We shall further see that it is possible to write the result in terms of familiar classical functions despite the complexity of the integrand in (\ref{seriesB}).

The basic results we need follow from simple changes of variable and subsequent algebraic manipulation and are (the definitions of the special functions used are given in the appendix)
\begin{align}
	\mathcal{L}\left[\left(1 - 4t\right)^{-\frac{1}{2}}\right](s) &=  \sqrt{\frac{\pi}{4s}} e^{-\frac{s}{4}}\left(i +\erfi{\sqrt{\frac{s}{4}}}\right) \nonumber \\
	\mathcal{L}\left[\left(1 + 4t\right)^{-\frac{1}{2}}\right](s) &=  \sqrt{\frac{\pi}{4s}} e^{\frac{s}{4}}\erfc{\sqrt{\frac{s}{4}}}
\end{align}
from which we use (\ref{Laplace}) to formally deduce the regulated parts of the Laplace transforms for $\Re(s) > 0$ up to constants to be fixed later:
\begin{align}
	\mathcal{L}\left[\left( \vphantom{\sqrt{1 - 4t}}t \sqrt{1 -4t} \right)^{-1} \right](s) &\simeq -\frac{s}{2}\pFq{2}{2}{\{1,1\}}{\{\frac{3}{2}, 2\}}{-\frac{s}{4}} \nonumber \\
	&- i\pi\erf{\sqrt{\frac{s}{4}}} \nonumber \\
	\mathcal{L}\left[\left( t \sqrt{1 + 4t} \right)^{-1}\right](s) 
	&\simeq \frac{s}{2} \pFq{2}{2}{\{1,1\}}{\{\frac{3}{2}, 2\}}{\frac{s}{4}} \nonumber \\
	&- \pi \erfi{\sqrt{\frac{s}{4}}} - i\pi \, ,
\end{align}
and similarly
\begin{align}
	\mathcal{L}\left[\left( \vphantom{\sqrt{1 - 4t}}t^{2}\sqrt{1 - \vphantom{+} 4t}\right)^{-1}\right](s) &\simeq  \frac{s^{2}}{4}\pFq{2}{2}{\{1,1\}}{\{\frac{3}{2}, 3\}}{-\frac{s}{4}} \nonumber \\ 
	&+ 2i \sqrt{\pi s}e^{-\frac{s}{4}} + i\pi\left(s-2\right) \erf{\sqrt{\frac{s}{4}}}\nonumber \\
	\mathcal{L}\left[\left(t^{2}\sqrt{1 + 4t}\right)^{-1}\right](s)	&\simeq  -\frac{s^{2}}{4} \pFq{2}{2}{\{1,1\}}{\{\frac{3}{2}, 3\}}{\frac{s}{4}}\nonumber \\
	&- 2\sqrt{\pi s}e^{\frac{s}{4}} + \pi\left(s+2\right) \erfi{\sqrt{\frac{s}{4}}} \nonumber \\
	&+i\pi s.
\end{align}
Putting these together, after suitable manipulations the integral in (\ref{seriesB}) can finally be written in terms of familiar functions as
\begin{align}
	&\int_{0}^{\infty}dt \left(\frac{2t -1}{4t^{2}\sqrt{1 - 4t}} + \frac{2t+1}{4t^{2}\sqrt{1 + 4t}}\right) e^{-s t}  \nonumber \\
	&=1 +  \frac{i\pi s}{4} - \frac{1}{2}\left(1+i\right)\sqrt{\pi s} \left(\cosh{\left(\frac{s}{4}\right)} - i \sinh{\left(\frac{s}{4}\right)}\right) \nonumber  \\
		&+ \bigg[ \frac{\pi s}{4} \erfi{\sqrt{\frac{s}{4}}} + \frac{s}{4} \pFq{2}{2}{\{1,1\}}{\{\frac{3}{2},2\}}{\frac{s}{4}} \nonumber \\
		&- \left(\frac{s}{4}\right)^{2} \pFq{2}{2}{\{1,1\}}{\{\frac{3}{2},3\}}{\frac{s}{4}} + s \longleftrightarrow \,-s \,\bigg] 
	\label{intRes}
\end{align}
In the above equation we have used the requirement that its limit as $z\rightarrow 0$ (or as $s \rightarrow \infty$) must vanish to determine the required (finite) constant piece that has hitherto been neglected. One may verify the absence of a divergence for positive real values of $s$ in the remaining terms. We have also carefully checked this analytic result against a numerical evaluation of the integral and found complete agreement.

Finally we may return to the commutator (\ref{CommS}) to investigate the consequences of our calculation. Using (\ref{intRes}) in (\ref{seriesB}) we find our main result:
	\begin{align}
	\left[r, \varphi\right]_{\star} &=  \pi r - \left(1-i\right)\sqrt{\pi \theta} \left(\cosh{\left(\frac{r^{2}}{\theta}\right)} - i \sinh{\left(\frac{r^{2}}{\theta}\right)}\right) \nonumber  \\
		&-i \bigg[ \pi r \erfi{\sqrt{\frac{r^{2}}{\theta}}} + r\, \pFq{2}{2}{\{1,1\}}{\{\frac{3}{2},2\}}{\frac{r^{2}}{\theta}} \nonumber \\
		&- \frac{r^{3}}{\theta}\, \pFq{2}{2}{\{1,1\}}{\{\frac{3}{2},3\}}{\frac{r^{2}}{\theta}} + (\theta, r) \leftrightarrow (-\theta, -r) \,\bigg],
	\label{intResTheta}
\end{align}
where as in (\ref{intRes}) we have used the $\leftrightarrow$ notation to signify that to the first three terms in square brackets should be added their value upon negation of $\theta$ and $r$. Inspection of (\ref{intResTheta}) reveals that a direct consequence of this resummation is that the commutator picks up a real part, a point to which we later return.

\subsection{Limiting cases}
\begin{figure}
	\includegraphics[width=0.4\textwidth]{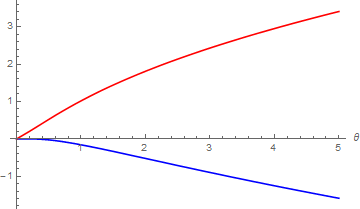}
	\caption{The real (blue line) and imaginary (red line) parts of the commutator as a function of $\theta$ for the illustrative case $r=1$.}
	\label{figTheta}
\end{figure}

\begin{figure}
	\includegraphics[width=0.4\textwidth]{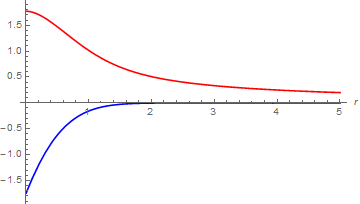}
		\caption{The real (blue line) and imaginary (red line) parts of the commutator as a function of $r$ for the illustrative case $\theta=1$.}
	\label{figR}
\end{figure}

Now that we have an analytic expression for the commutator, we may explore its behaviour in different regimes. We first use our result to revisit the limiting case of (pseudo-)commutativity, whereby $\theta \rightarrow 0$ or $r \rightarrow \infty$. In this case we have found (see the series expansions for the relevant functions in the Appendix) that the commutator has expansion
\begin{equation}
	\left[r, \varphi\right]_{\star} = \frac{i\theta}{r} + \frac{i \theta^{3}}{4r^{5}} + \mathcal{O}\left(\frac{\theta^{5}}{r^{9}}\right),
	\label{small}
\end{equation} 
which correctly reproduces the original (purely imaginary) series (\ref{series}) term by term. It is worth recalling that the apparent small-$r$ divergence here causes no problems for the analysis of black holes in \cite{btz} since attention is limited to coordinates $r > 0$. At the other extreme we can investigate the large $\theta$ or small $r$ limit in which $z \rightarrow \infty$. Then we have shown that the commutator has leading order behaviour
\begin{equation}
	\left[r, \varphi\right]_{\star} = -\left(1-i\right)\sqrt{\pi\theta} + \pi r + \mathcal{O}\left(\frac{r^{2}}{\sqrt{\theta}}\right)
	\label{large}
\end{equation}
with further corrections in powers of $\frac{r^{2}}{\sqrt{\theta}}$ whose coefficients decrease faster than the inverse square of their order. This suggests that when $r^{2}$ is small in comparison to $\theta$ the commutator is approximately constant with respect to radial distance, which warrants further investigation as we discuss below. The complete analytic result (\ref{intResTheta}) allows us to interpolate between these two limits, although its complexity makes a plot of its shape for different values of $\theta$ and $r$ more illuminating. To this end, Figures \ref{figTheta} and \ref{figR} show the real and imaginary parts of the commutator for suitable values of the parameters $\theta$ and $r$.

We have also determined that if we had instead chosen the contour of integration in (\ref{intRes}) to run just below the square root branch cut on the real axis then the only change is to the imaginary part of the result, whose sign is directly reversed. Since both choices lead to valid resummations of (\ref{series}), and there is no underlying field theory from which to derive a physical reason to favour one choice over another, one may average these two solutions so as to arrive at a real-valued function; this would consequently allow us to recover a purely imaginary commutator, desirable if the coordinate operators are to be Hermitian as is usual. In this case, our large $\theta$ expansion becomes
\begin{align}
	\left[r, \varphi\right]_{\star} &= i\sqrt{\pi \theta}  -i \sqrt{\frac{\pi}{\theta}}r^{2} +\ldots &&\theta \rightarrow \infty
	\label{limits}
\end{align}
whilst (\ref{small}) remains unchanged.
\section{Conclusion}
We have considered the fundamental commutation relations for a two-dimensional non-commutative space described by polar coordinates $\hr$ and $\hphi$. By applying the Borel resummation technique we have defined and evaluated the power series expression for $\left[r, \varphi\right]_{\star}$, verifying the analytic result agrees with a numerical calculation. Our answer can be written in terms of functions that are well-known to the mathematics community. Furthermore, we have shown that our result is consistent with the commutator previously proposed in the literature at lowest order in the non-commutativity parameter $\theta$. The advantage of our analytic calculation is that it can now be extended to arbitrary values of $\theta$ and radial distance and we have in particular considered the regime where $r^{2} \ll \theta$, finding the commutator's leading order behaviour to depend only upon (the square root of) $\theta$. 

One immediate application of our result is to the commutator $\left[\hr^{2}, \hphi\right]$ mentioned in the introduction. In the current context, (\ref{small}) implies corrections to this at higher order in $\theta$,
\begin{equation}
	\left[r^{2}, \varphi\right]_{\star} = 2i \theta + \frac{i \theta^{3}}{2r^{4}} + \ldots,
	\label{r2t}
\end{equation}
and when $\theta$ is large in comparison to $r^{2}$ (\ref{limits}) provides
\begin{equation}
	\left[r^{2}, \varphi\right]_{\star} = 2i \sqrt{\pi \theta} r - 2i\sqrt{\frac{\pi}{\theta}}r^{3} + \ldots,
	\label{r2t2}
\end{equation}
whereas in \cite{andy}, Iskauskas found the series expansion arising from application of (\ref{star}) terminates at the $\mathcal{O}(\theta)$ contribution in (\ref{r2t}). For this reason it would be interesting to investigate the consequences of the higher order terms of (\ref{r2t}) in the context of fuzzy BTZ black holes. In this setting, it may be the case that some consistency condition can be used to determine which form of the commutator is physical meaningful.

Furthermore, it would be worthwhile also using (\ref{large}), or its cousin in (\ref{limits}), to probe the large $\theta$ limit of these objects. Indeed, the the spatial dependence of the fundamental commutation relations in this limit is easily deduced by determining the Cartesian commutator $\left[x, y\right]_{\star}$ in this extreme limit. This can be achieved either by using (\ref{star}), based on (\ref{limits}), with $f(r, \theta) = r \cos{\left(\theta\right)}$ and $g(r, \theta) = r\sin{\left(\theta\right)}$ or by direct calculation of $\left[\hr \cos{(\hphi)}, \hr \sin{(\hphi)}\right]$ from (\ref{limits}). In either case it is straightforward to obtain
\begin{equation}
	\left[x, y\right]_{\star} 
	\propto i\sqrt{\theta}\sqrt{x^{2} + y^{2}} ,
	\label{xyr}
\end{equation}
which shows that the non-commutativity of the coordinates becomes position dependent in this limit. A representation of this algebra is easily determined by noting that the relation (\ref{limits}) is at leading order solved by operators satisfying the Heisenberg algebra, which can subsequently be transformed to coordinates fulfilling (\ref{xyr}). One example is to parameterise the two-dimensional space by the eigenvalues of the angular coordinate, $\hphi$, and take $\hx = i\sqrt{\theta}\partial_{\varphi}\cos{(\varphi)}$ and $\hy = i\sqrt{\theta}\partial_{\varphi}\sin{(\varphi)}$, whereby $\hx^{2} + \hy^{2} = -\theta\partial^{2}_{\varphi}$. We intend to investigate black hole solutions in this curious limit of high non-commutativity to complement previous work carried out in this area. It is clear that many open questions remain on the subject of non-commutativity in polar coordinates and we hope that this article makes a positive contribution towards research in this area.

\section{Acknowledgements}
The author is grateful for constructive comments on the manuscript from Paul Cook, Andrew Iskauskas and Daniele Galloni. Sincere and warm thanks also to Jorge Bruno,  Nadia Jaszczynska and Marcus Kaiser for helpful discussions, recommending various resources and encouragement throughout this work, without which support this article would not have come to be. 
\appendix
\section{Special functions}
We list here the definitions of the special functions used in the main text so as to provide a complete reference for the calculations described therein ($x$ is a generally complex argument to these functions):
\begin{align*}
	\erf{x} &\equiv \sqrt{\frac{4}{\pi}} \int_{0}^{x}e^{-y^{2}} dy \\
	\erfc{x} &\equiv 1 - \erf{x} \\
	\erfi{x} &\equiv -i\erf{ix} \\
	\Gamma\left[x\right] &\equiv \int_{0}^{\infty} y^{x-1}e^{-y} dy \\
	\Gamma\left[x, \alpha\right] &\equiv \int_{\alpha}^{\infty} y^{x -1} e^{-y} dy \\
	\pFq{p}{q}{\{a_{i}\}}{\{b_{j}\}}{x} &\equiv \sum_{n=0}^{\infty} \frac{(a_{1})_{n}\ldots (a_{p})_{n}}{(b_{1})_{n}\ldots (b_{q})_{n}} \frac{x^{n}}{n!}.
\end{align*}
In the final definition of the generalised hypergeometric function we have made use of the Pochhammer symbol, defined by $(\beta)_{0} =1$ and
\begin{equation}
	(\beta)_{n} = \beta(\beta + 1)\cdots (\beta + n - 1)= \frac{\Gamma[\beta + n]}{\Gamma[\beta]}
\end{equation}
for $n > 0$. A useful property for the results in the main text is (for $x > 0$) $\sqrt{\pi} \Gamma\left[\frac{1}{2}, -x\right] = \pi\left(1 - i \erfi{\sqrt{x}\,}\right)$.
\subsection{Series expansions}
For reference we include here the leading order series expansions of the functions that enter the main result. In the text these were employed to determine the limiting behaviour of the polar commutator. We first consider the series expansions for small arguments:
\begin{align*}
	\erfi{x} &= \frac{2x}{\sqrt{\pi}} + \frac{2x^{3}}{3\sqrt{\pi}} + \mathcal{O}\left(x^{5}\right) \\
	\pFq{2}{2}{\{1,1\}} {\{\frac{3}{2}, 2\}}{x}&= 1+ \frac{x}{3}+\frac{4x^{2}}{45}+\frac{2x^{3}}{105} + \mathcal{O}\left(x^{4}\right) \\
	\pFq{2}{2}{\{1,1\}}{ \{\frac{3}{2}, 3\}}{x}&= 1+ \frac{2x}{9}+\frac{2x^{2}}{45}+\frac{4x^{3}}{525} + \mathcal{O}\left(x^{4}\right).
\end{align*}
Furthermore, taking the expansion about $x \rightarrow \pm \infty$ we find (we denote by $\Theta(x)$ the Heaviside step function):
\begin{align*}
	\erfi{\pm x} &= \mp i \pm e^{x^{2}}\left(\frac{1}{\sqrt{\pi} x}+ \mathcal{O}\left(x^{-3}\right)\right) \\
	\pFq{2}{2}{\{1,1\}} {\{\frac{3}{2}, 2\}}{\pm x}&= \left(-i\pi\Theta(x) \mp \ln{x} \pm \psi\left(\frac{1}{2}\right)\right)\frac{1}{2x} \\
	& + i^{1-\Theta(x)}e^{\pm x}\left(\frac{\sqrt{\pi}}{2x^{\frac{3}{2}}} + \mathcal{O}\left(x^{-\frac{5}{2}}\right)\right) \\
	& + \mathcal{O}\left(x^{-2}\right)\\
	\pFq{2}{2}{\{1,1\}} {\{\frac{3}{2}, 3\}}{\pm x}&= \left( \pm 1-i\pi\Theta(x) \mp \ln{x} \pm \psi\left(\frac{1}{2}\right)\right)\frac{1}{x} \\
	&- \left(1+i\pi\Theta(x) +\ln{x} - \psi\left(-\frac{1}{2}\right) \right)\frac{1}{2x^{2}} \\
	&+i^{1-\Theta(x)}e^{\pm x}\left(\pm\frac{\sqrt{\pi}}{x^{\frac{5}{2}}} + \mathcal{O}\left(x^{-\frac{7}{2}}\right)\right)\\
	&+ \mathcal{O}\left(x^{-3}\right),
\end{align*}
where $\psi(x) \equiv \left(\ln\left(\Gamma[x]\right)\right)^{\prime}$ is the digamma function. We have expanded the last function to next to leading order which is required due to its prefactor of $\frac{r^{3}}{\theta}$ in the main result (\ref{intResTheta}). These series are sufficient to verify the limits discussed in the article.
\section*{References}
\bibliographystyle{unsrt}
\bibliography{bibPol}
\end{document}